\documentclass{emulateapj}

\usepackage{amsmath}
\usepackage{txfonts}
\usepackage{natbib}
\usepackage{graphicx}

\shorttitle{Neutron star kicks in three dimensions}

\shortauthors{Wongwathanarat, Janka, \& M\"uller}

\begin{document}

\title{Hydrodynamical neutron star kicks in three dimensions}

\author{Annop Wongwathanarat,
        Hans-Thomas Janka,
        and 
        Ewald M\"uller
       }
\affil{Max-Planck-Institut f\"ur Astrophysik,
       Karl-Schwarzschild-Str. 1, D-85748 Garching, Germany;
       annop@mpa-garching.mpg.de}

\begin{abstract}
Using three-dimensional (3D) simulations of neutrino-powered 
supernova explosions we show that the hydrodynamical kick 
scenario proposed by Scheck et al.\ on the basis of two-dimensional
(2D) models can yield large neutron star (NS) recoil velocities also
in 3D. Although the shock stays relatively spherical, standing
accretion-shock and convective instabilities lead to a globally
asymmetric mass and energy distribution in the postshock layer.
An anisotropic momentum distribution of the ejecta is built up only 
after the explosion sets in. Total momentum conservation implies the 
acceleration of the NS on a timescale of 1--3 seconds, mediated mainly
by long-lasting, asymmetric accretion downdrafts and the anisotropic 
gravitational pull of large inhomogeneities in the ejecta. In a 
limited set of 15\,$M_\odot$ models with an explosion energy of 
about 10$^{51}$\,erg this stochastic mechanism is found to
produce kicks from $<$100\,km\,s$^{-1}$ to $\ga$500\,km\,s$^{-1}$, 
and $\ga$1000 km\,s$^{-1}$ 
seem possible. Strong rotational flows around the accreting NS do
not develop in our collapsing, non-rotating progenitors. The NS 
spins therefore remain low with estimated periods of
$\sim$500--1000$\,$ms and no alignment with the kicks.
\end{abstract}


\section{Introduction}

Young neutron stars (NSs) possess average space velocities around 
400\,km\,s$^{-1}$, much larger than those of their progenitor stars, 
implying that they are accelerated during the birth in a supernova (SN)
explosion \citep[e.g.,][]{FaucherKaspi06,Hobbsetal05,Arzoumanianetal02}.
Moreover, alignment of the NS spin and kick
was inferred for the Crab and Vela pulsars \citep{Kaplanetal08,
NgRomani07} and several other young pulsars 
\citep[see][and references therein]{Wangetal06} from comparisons 
of the direction of proper motion with the projected rotation axis
as determined from the symmetry axis of the pulsar wind nebula on
X-ray images. The same conclusion was drawn for samples of radio
pulsars from the linear polarization of the pulses, whose position
angle reflects the spin direction \citep{Johnstonetal05,Rankin07}.
However, the Crab and Vela pulsars may not be good cases for determining
misalignments because they are moving at smaller speeds than the
average pulsar population and therefore the unknown velocity of the
progenitor implies a bigger uncertainty. On the other hand, also the radio
data do not seem to make a clear case for a general alignment of spin
and kick directions in the reference frame of the progenitor's motion
\citep{Johnstonetal07}.

Analysing the observational information, also of characteristics of
NS binaries, \citet{Laietal01} concluded that the NSs
received their kicks most probably at the time of the SN explosion.
A large variety of mechanisms for natal kicks has been proposed,
either by hydrodynamical effects linked to large-scale asymmetries
of the SN explosion \citep[e.g.,][]{Herant95,BurrowsHayes96,
JankaMueller94,Thompson00,Schecketal04,Schecketal06}
or by anisotropic neutrino emission from the nascent NS
\citep[e.g.,][]{Chugai84,BurrowsWoosley86,Socratesetal05}.
However, it is very difficult to produce even only a one-percent
global dipole asymmetry of the neutrino emission, which is needed for
a kick of 300\,km\,s$^{-1}$. For this to be possible one has to invoke
controversial assumptions like very strong global dipolar magnetic 
fields inside the NS \citep[$\ga 10^{16}$\,G; e.g.,][]{ArrasLai99},
arguable neutrino properties \citep[e.g., sterile neutrinos, large 
neutrino magnetic moments; e.g.,][]{Fulleretal03},
or unsettled mechanisms to create strong emission asymmetries
in the neutrinospheric region \citep[e.g.,][]{Socratesetal05}.

On the basis of 2D SN models \citet{Schecketal04,Schecketal06} 
argued that the standing accretion shock instability 
\citep[SASI;][]{Blondinetal03,FoglizzoTagger00,Foglizzo02}, 
which grows after shock
stagnation and causes large global non-radial asymmetry of the 
accretion flow to the NS and of the beginning SN explosion, can lead
to kicks of typically several hundred km\,s$^{-1}$ and even more than
1000\,km\,s$^{-1}$ if the dipole ($\ell = 1$) component of the asymmetry
was sufficiently strong. \citet{BlondinMezzacappa07} showed in
idealized, stationary-accretion setups that SASI spiral modes may
also have the potential to generate pulsar spin periods consistent 
with observations. In this {\em Letter} we present the first
simulations of SN explosions from core bounce to $\sim$1.5\,s 
later that confirm the potential of SASI-induced asymmetries to 
produce typical NS kicks in the more realistic 3D environment of
collapsing stellar cores.

\section{Numerical setup}

We use the explicit finite-volume, Eulerian, multi-fluid hydrodynamics
code {\sc Prometheus} \citep{PROMET1,PROMET2,PROMET3}.
The advection of nuclear species is treated by the Consistent
Multi-fluid Advection (CMA) scheme of \citet{CMA}, and self-gravity 
according to \citet{Gravity}. 
General relativistic corrections of the monopole are taken into
account as in \citet{Schecketal06} and \citet{Arconesetal07}.

All simulations are computed on an axis-free overlapping 
``Yin-Yang'' grid \citep{YinYang} in spherical polar coordinates,
which was recently implemented into our code \citep{YYmethod}. We use
$400(r)\times47(\theta)\times137(\phi)\times2$ grid cells 
corresponding to an angular resolution of $2^\circ$ and covering the
full $4\pi$ solid angle. Better resolution in the inner region of our
computational domain is achieved by a constant radial zone size of 
about 0.030\,km up to $r \approx 100$\,km 
The radial grid 
is logarithmically spaced beyond this radius up to an outer grid
boundary of $R_\mathrm{ob} = 18000$\,km, which is sufficient to prevent 
the SN shock from leaving the computational domain during the
simulated time. At $R_\mathrm{ob}$ a free outflow boundary is chosen.

The high-density inner core of the proto-NS (PNS) is excised and 
replaced by a point mass at the coordinate origin. The shrinking of
the PNS is mimicked by a retraction of the closed inner boundary at 
$R_\mathrm{ib}$ together with the radial grid. Hydrostatic equilibrium 
is assumed at $R_\mathrm{ib}$. The stellar fluid is described by the
tabulated microphysical equation of state (EoS) of \citet{JankaMueller96}. 
Neutrino transport and neutrino-matter interactions are
approximated as in \citet{Schecketal06} by radial integration of the 
one-dimensional (spherical), grey transport equation for all angular 
grid directions ($\theta$,\,$\phi$) independently. This ``ray-by-ray'' 
approach allows for angular variations of the neutrino fluxes.

%------------------------------------------------------------------------
\begin{table}
\caption{Explosion and NS properties for all models at 1.4\,s
after bounce.}
\begin{center}
\begin{tabular}{lcccccccc}
\hline
\hline
Model & $M_\mathrm{ns}$ & $t_\mathrm{exp}$ & $E_\mathrm{exp}$ &
$v_\mathrm{ns}$ & $a_\mathrm{ns}$ & $J_{\mathrm{ns},46}$ &
$\alpha_\mathrm{sk}$ & $T_\mathrm{spin}$\\
      & [$M_\odot$] & [ms] & [B] &
[km/s] & [km/s$^2$] & [g\,cm$^2$/s] & [$^\circ$] & [ms]\\
\hline
W15-1 & 1.37 & 246 & 1.12 & 331 & 175 & 1.51 & 117 & 652\\
W15-2 & 1.37 & 248 & 1.13 & 405 & 144 & 1.56 & 58  & 632\\
L15-1 & 1.58 & 421 & 1.13 & 161 & 66  & 1.89 & 148 & 604\\
L15-2 & 1.51 & 381 & 1.74 & 78  & 3   & 1.04 & 62  & 1041\\
\hline
\end{tabular}
\end{center}
\label{tab:results}
\end{table}
%------------------------------------------------------------------------

%-----------------------------------------------------------------------
\begin{figure}
\plotone{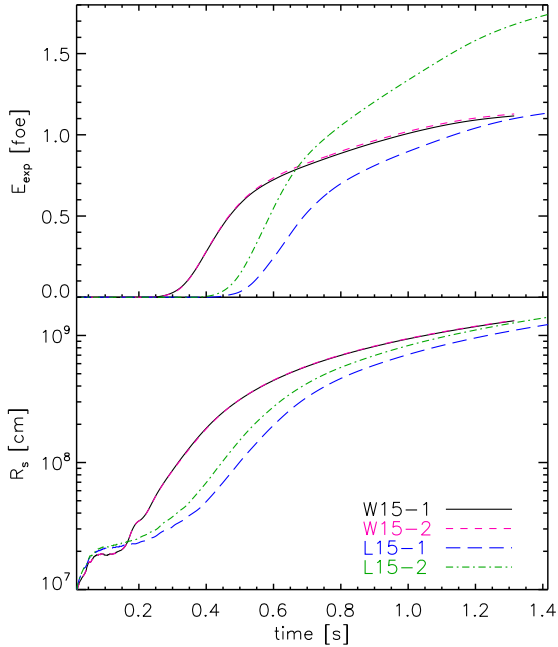}
\caption{Time evolution of the SN explosion energy $E_\mathrm{exp}$
  ({\em top}) and of the average shock radius $R_\mathrm{s}$ for all
  computed models. Since models W15-1 and W15-2 differ only in the initial
  random seed perturbations of the radial velocity field, the results
  for these models are essentially identical and the short-dashed and
  solid lines lie on top of each other. Despite the similarity of the
  global parameters, both models develop different explosion asymmetries
  and neutron star kicks (see Table~\ref{tab:results} and
  Fig.~\ref{fig:NSevolution}).}
\label{fig:SNevolution}
\end{figure}
%------------------------------------------------------------------------

\section{Investigated models}

Our models W15 and L15 are based on two non-rotating 15\,$M_\odot$ 
progenitors, s15s7b2 of \citet{WoosleyWeaver95} and a star evolved by
\citet{Limongi2000}, which were followed through collapse to 15\,ms 
after bounce with the
{\sc Prometheus-Vertex} code in one dimension (R.~Buras and A.~Marek, 
private communication). To break spherical symmetry, random seed 
perturbations of 0.1\% are imposed on the radial velocity ($v_r$)
field. Explosions  
with chosen energy are initiated by neutrino heating depending
on suitable values of the neutrino luminosities imposed at
the lower boundary.

Our models W15-1 and W15-2 differ only by the initial seed perturbations,
while L15-2 is set to have a higher explosion energy than L15-1.
Table~\ref{tab:results} lists the corresponding explosion energies
$E_\mathrm{exp}$ at $\sim$1.4\,s p.b.\ and explosion times $t_\mathrm{exp}$
defined as the instant when $E_\mathrm{exp}$ (i.e., the total energy of all
zones where the internal plus kinetic plus gravitational energy is $>$0)
is 10$^{48}$\,erg. 
Figure~\ref{fig:SNevolution} shows that this moment coincides
roughly with the time when the average SN shock radius,
$R_\mathrm{s}$, exceeds 500\,km.
The W15 models explode earlier and their NS (baryonic) mass,
$M_\mathrm{ns}$ (defined as enclosed mass at a radius
$R_\mathrm{ns}$ where the density is $10^{11}$\,g\,cm$^{-3}$),
is therefore smaller.

%-----------------------------------------------------------------------
\begin{figure*}
\plotone{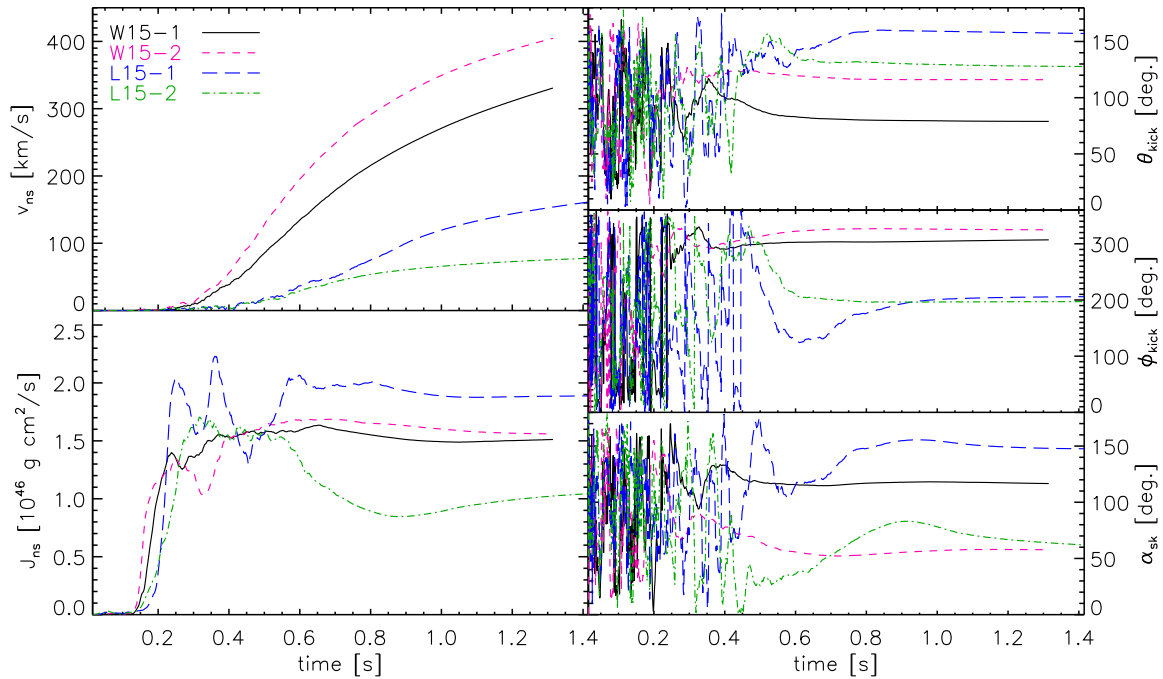}
\caption{Time evolution of NS velocity $v_\mathrm{ns}$
  ({\em top-left}), angular momentum $J_\mathrm{ns}$ ({\em bottom-left}),
  latitudinal ({\em top-right}) and azimuthal ({\em middle-right}) angles
  of the NS kick vector, $\theta_\mathrm{kick}$ and $\phi_\mathrm{kick}$,
  respectively, and the angle $\alpha_\mathrm{sk}$ between the NS spin
  and kick directions ({\em bottom-right}).}
\label{fig:NSevolution}
\end{figure*}
%------------------------------------------------------------------------

\section{Neutron star kicks and spins}

Neutrino heating around the NS triggers delayed explosions, which
are preceded (and supported) by a postbounce phase of several 100\,ms
of violent SASI and convective activity. This leads to large-scale
asymmetries of the ejecta, potentially imposing a kick and spin to
the NS, for which we evaluate our simulations in a post-processing 
step. The time evolution of the NS kick velocity $v_\mathrm{ns}$,
NS angular momentum $J_\mathrm{ns}$, lateral and longitudinal
angles of the kick direction ($\theta_\mathrm{kick}$ and 
$\phi_\mathrm{kick}$, respectively), and the angle 
$\alpha_\mathrm{sk}$ between spin and kick directions are plotted 
in Fig.~\ref{fig:NSevolution}. The final numbers are given in
Table~\ref{tab:results} ($J_\mathrm{ns}$ values there are normalized
by 10$^{46}$), supplemented by the NS acceleration 
$a_\mathrm{ns}$ at the end of the runs and our estimated NS spin
periods $T_\mathrm{spin}$.

\begin{figure*}
\resizebox{0.33\textwidth}{!}{
%\includegraphics*[0.5cm,4.7cm][20cm,21cm]
%   {fig2-1.ps}}
\includegraphics*{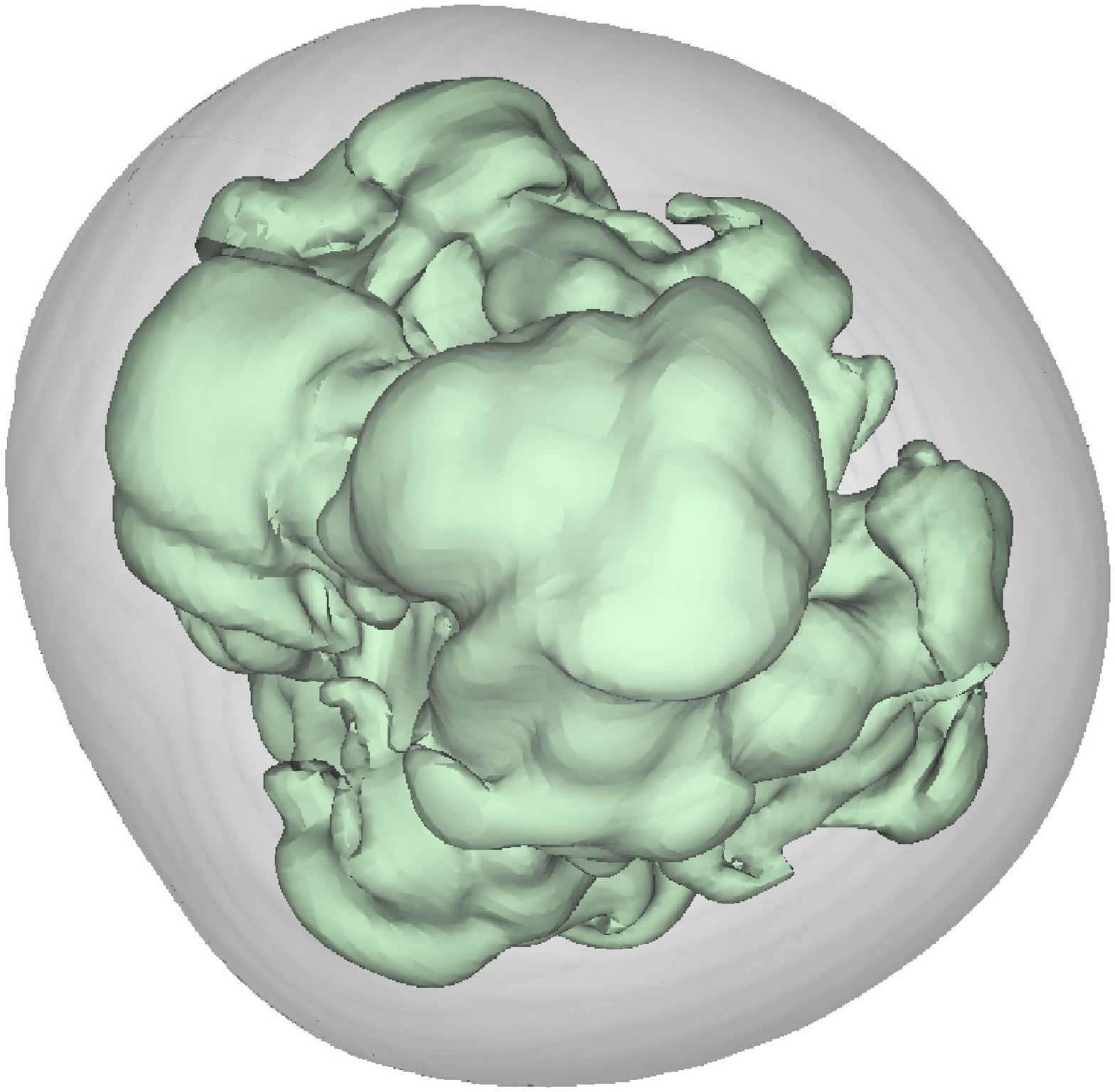}}
\resizebox{0.33\textwidth}{!}{
\includegraphics*{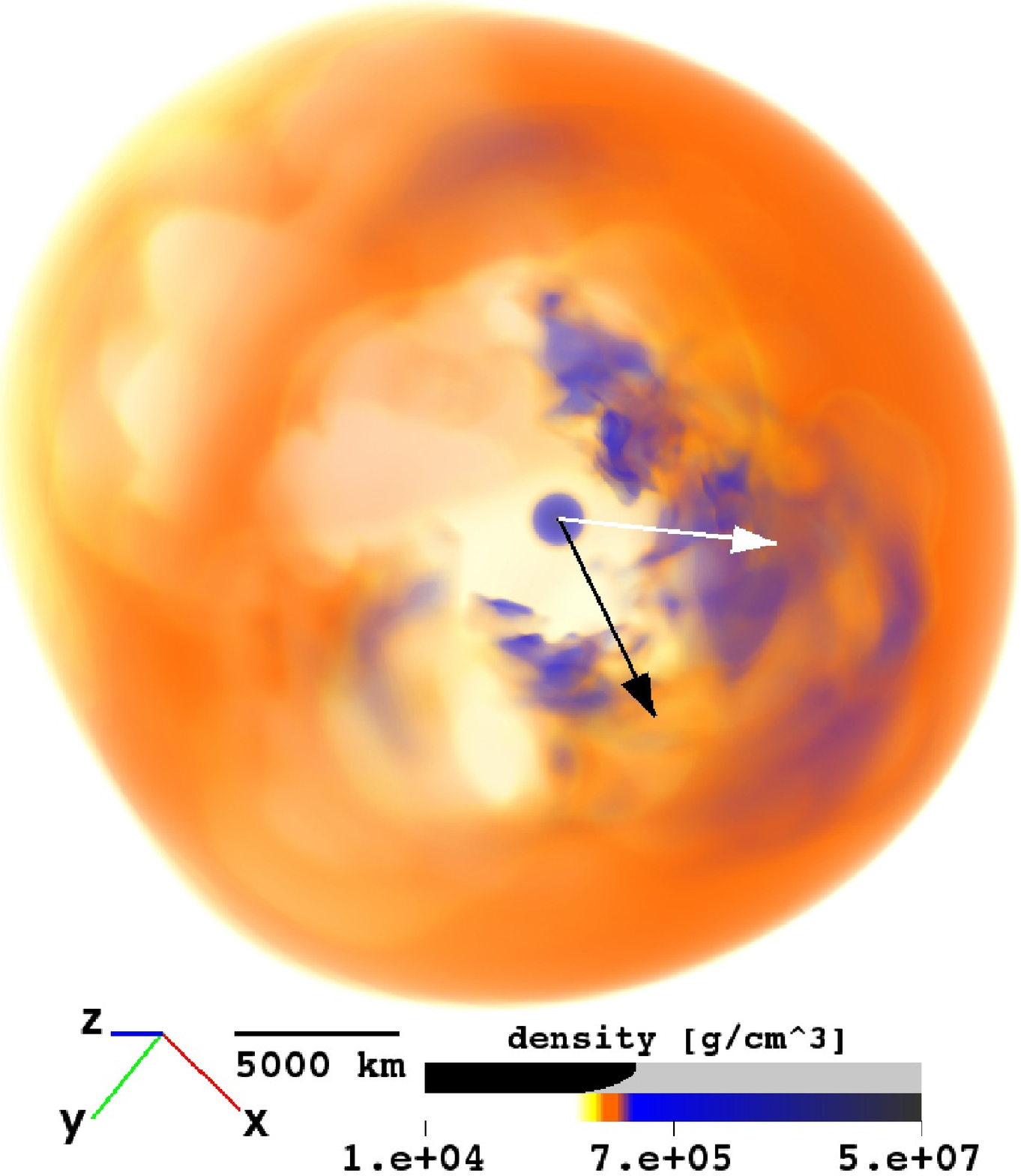}}
\resizebox{0.33\textwidth}{!}{
%\includegraphics*[0.5cm,4.7cm][20cm,21cm]
%   {fig2-2.ps}}\\
\includegraphics*{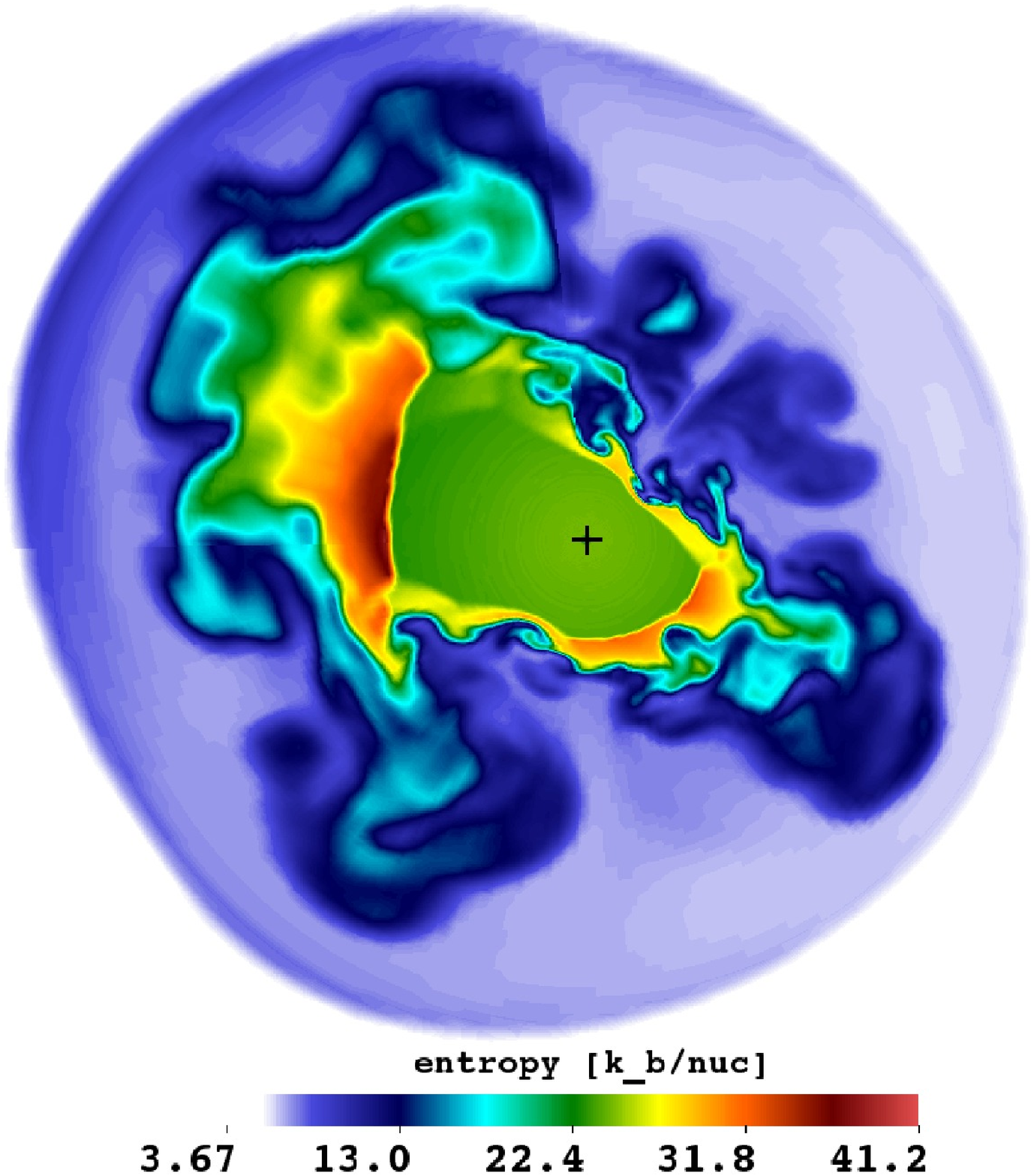}}\\
\caption{Entropy-isosurfaces of the SN shock and the high-entropy
  bubbles ({\em left}), ray-casting image of the density ({\em middle}),
  and entropy distribution in a cross-sectional plane through the center
  ({\em right}) at $t=1.3$\,s after bounce for model W15-2. The outer
  boundaries coincide with the shock surface, the viewing
  direction is normal to the plane of the NS kick and spin vectors, which
  also define the plane for the entropy slice. The kick and spin directions
  are indicated by the white and black arrows, respectively, in the
  {\em middle figure}.
  The NS location is also marked by a black cross in the {\em right plot}
  and is clearly displaced from the geometrical center of the expanding shock.
  The SN shock has an average radius of 13000\,km (a length of 5000\,km is
  given by a yardstick below the {\em middle image}) but shows a pronounced
  dipolar deformation, which is clearly visible from the color asymmetry
  of the postshock gas between the lower right (weaker shock with minimum
  radius of 11000\,km) and upper left (stronger shock with maximum radius
  of 15000\,km) directions. The {\em middle plot} corresponds roughly
  to the projection of the density distribution on the observational
  plane. Dilute bubble regions are light-colored in white and yellow,
  while dense clumps appear more intense in reddish and bluish hues.
  The blue circle around the NS represents the dense inner region
  of the spherically symmetric neutrino-driven wind. This wind is
  visible in green in the {\em right image} and is bounded by the
  aspherical wind termination shock. The wind is shocked to higher 
  entropies on the left side, where it passes the termination shock
  at larger radii because of the faster expansion of the preceding 
  SN ejecta.}
\label{fig:W15-2-3D}
\end{figure*}

Since the NS is excised and replaced by
a point mass at the grid center, it cannot travel. Nevertheless,
like a wall reflecting a bouncing ball, it can absorb momentum
(but will not move as if it had an infinite inertial mass). 
Because of total linear momentum conservation in 
the rest frame of the progenitor, we can 
\citep[following][]{Schecketal06}
compute $\mathbf{v}_\mathrm{ns}$ from the negative of
the total momentum $\mathbf{P}_\mathrm{gas}$ of the gas outside
of the NS as
\begin{equation}
\mathbf{v}_\mathrm{ns}(t)=-\mathbf{P}_\mathrm{gas}(t)/M_\mathrm{ns}(t)
\, ,
\label{eq:vns}
\end{equation}
where $\mathbf{P}_\mathrm{gas}=\int_{R_\mathrm{ns}}^{R_\mathrm{ob}}
\mathrm{d}V\,\rho\mathbf{v}$. Tests confirmed very good linear
momentum conservation of our code, while angular momentum is more
difficult to conserve, e.g., when a rotating gas mass is in rapid 
motion across large distances on the grid. We therefore estimate
$\mathbf{J}_\mathrm{ns}$ again as the negative
of the angular momentum of the exterior gas, but by integrating only
over the volume between $R_\mathrm{ns}$ and 
$r_\mathrm{o} = 500$\,km and adding to this
the angular momentum that is carried by the gas flux leaving this  
sphere:
\begin{equation}
\mathbf{J}_\mathrm{ns}(t)=-\left(\int_{R_\mathrm{ns}}^{r_\mathrm{o}}
\mathrm{d}V \,\rho \mathbf{j}(t) + A_\mathrm{o} \int_0^t
\mathrm{d}t' \, (\rho \mathbf{j}v_r)\mid_{r_\mathrm{o}}\right) \, ,
\label{eq:Jns}
\end{equation}
where $\mathbf{j}$ is the specific angular momentum
and $A_\mathrm{o} =4\pi r_\mathrm{o}^2$. 
This assumes that the asymmetric gas mass outside of $r_\mathrm{o}$
does not exert any important torque on the gas mass below 
$r_\mathrm{o}$. Accordingly, we see $\mathbf{J}_\mathrm{ns}(t)$ 
asymptoting (see Fig.~\ref{fig:NSevolution}) when accretion on
the NS ends (around 0.6--0.8\,s after bounce). At this time
%   (in W15-1 at $\sim$0.7\,s, in W15-2 and L15-2 at 
%   $\sim$0.6\,s, and in L15-1 at $\sim$0.8\,s). 
the asymmetric downdrafts of cool gas filling the volumes between 
rising bubbles of high-entropy, neutrino-heated matter
(Fig.~\ref{fig:W15-2-3D}, left and right panels) do not reach
down below 500\,km, but are replaced by the spherically symmetric
neutrino-driven wind around the NS (green region in the right panel
of Fig.~\ref{fig:W15-2-3D}). 
Assuming $J_\mathrm{ns} = |\mathbf{J}_\mathrm{ns}| = \mathrm{const}$
after the end of our simulations (at $t \approx 1.4$\,s p.b.),
we obtain a rough estimate of the final NS spin period from
$T_\mathrm{spin} = 2\pi I_\mathrm{ns}/J_\mathrm{ns}$ by considering
a rigidly rotating, homogeneous sphere of mass $M_\mathrm{ns}$,
i.e., $I_\mathrm{ns}=\frac{2}{5} M_\mathrm{ns}R_\mathrm{ns}^2$,
with a final radius of $R_\mathrm{ns} = 12$\,km.

We find NS kick velocities between 80\,km\,s$^{-1}$ (L15-2) and
405\,km\,s$^{-1}$ (W15-2). In the latter case the acceleration at
1.3\,s p.b.\ is still about 150\,km\,s$^{-2}$ (Table~\ref{tab:results}).
Assuming that half of this value applies for another second leads
to a final kick of nearly 500\,km\,s$^{-1}$. Comparing 
Figs.~\ref{fig:SNevolution} and \ref{fig:NSevolution} one sees 
that the kick remains very small ($\la 10$\,km\,s$^{-1}$) and the 
kick direction fluctuates chaotically in all cases
before the explosion takes off. Only after the radial expansion
of the ejecta sets in and the asymmetry pattern of energy
and density distribution gets frozen in, the NS acceleration 
takes place and continues for 1--3 seconds \citep{Schecketal06}.
Accordingly, the kick finds its final direction only
shortly afterwards (at 0.4--0.5\,s in W15-1 and W15-2 and
0.6--0.8\,s in L15-1 and L15-2). 

Decomposing the radial integral 
$D(\theta,\phi) = \int_{R_\mathrm{ns}}^{R_\mathrm{s}}
\mathrm{d}r\,\rho(r)$ in spherical harmonics with the polar axis
chosen aligned with the kick vector, we find that the $\ell = 1$
mode clearly dominates after 0.3--0.4\,s p.b.\ in W15-1 and W15-2, 
the models with the largest kick, while this effect is absent
in the low-kick models L15-1 and L15-2. In Fig.~\ref{fig:W15-2-3D},
middle panel, the dipolar asymmetry of the explosion is clearly
visible. The NS recoil (white arrow) is opposite to the direction
of the fastest shock expansion (towards the upper left) driven by big
high-entropy bubbles. The lower density there is contrasted
by much higher density and slower gas expansion in the opposite direction. 
Accordingly, the geometrical center of the shock and ejecta shell
is displaced from the grid center (and NS location) by several
1000\,km (see middle and right panels of Fig.~\ref{fig:W15-2-3D}).

Besides kicks the PNSs attain angular momentum
(1--2$\times 10^{46}$\,g\,cm$^2$s$^{-1}$) without any obvious correlation 
in magnitude and direction to the recoil. We estimate final NS spin
periods between 600\,ms and 1000\,ms (Table~\ref{tab:results}). 
Evaluating $D(\theta,\phi)$ for spherical harmonics with the axis
parallel to the spin vector, an $\ell = 1,\,m = 1$ mode appears
as the dominant non-axisymmetric mode in the W15-models
after 0.4\,s p.b., reflecting the mass imbalance on both sides of
the spin vector as directly visible for W15-2 in the middle panel 
of Fig.~\ref{fig:W15-2-3D}.

%------------------------------------------------------------------------
%------------------------------------------------------------------------

\section{Physical origin of the NS kicks and spins}

The described hydrodynamic NS acceleration proceeds in three stages.
(1) Initially convective mass flows and SASI sloshing
motion of the postshock layer create an anisotropy of the 
mass-energy distribution around the PNS. Convective downdrafts,
channelling gas accreted through the stalled shock into the 
neutrino-heating region, get deflected 
to feed an asymmetric pattern of high-entropy 
bubbles. The energy-loaded bubbles are created, collapse again, 
and reappear in a quasi-chaotic way to become smaller or larger,
absorbing less or more neutrino energy. This stochastic bubble
formation, however, does not cause an appreciable recoil of 
the NS (Fig.~\ref{fig:NSevolution}). (2) When the explosion
sets in, the shock and postshock gas begin to expand aspherically,
driven by the asymmetric inflation of the bubbles. The ejecta gas
therefore gains radial momentum and its c.o.m.\ begins to 
shift away from the coordinate origin (Fig.~\ref{fig:W15-2-3D}):
The ejecta shell acquires a net linear momentum because of
different strengths of the explosion in different directions.
The initial {\em energy and mass asymmetry} is thus converted
to a {\em momentum asymmetry} by the conversion of thermal to
kinetic energy through hydrodynamical forces. When the 
expansion timescale becomes shorter than the timescale of lateral 
mixing, the asymmetric ejecta structures freeze in.
(3) Because of linear momentum conservation, the NS must receive
the negative of the total momentum of the anisotropically 
expanding ejecta mass. Hydrodynamic pressure forces alone cannot 
achieve the NS acceleration \citep{Schecketal06}. As long as accretion
downdrafts reach the NS, momentum is transferred by asymmetric gas 
flows. Stronger accretion on the weaker side of the blast and more
mass loss in the neutrino-driven wind on the other side cause
a recoil opposite to the main explosion direction. However,
the largest (time-integrated) kick contribution,
which continues even after accretion
has ended and the wind has become spherical (Fig.~\ref{fig:NSevolution}), 
results from the gravitational pull of the anisotropic
shell of ejecta \citep{Schecketal06}.
A hemispheric asymmetry of the mass distribution of 
$\Delta m = \pm 10^{-3}\,M_\odot$ in a shell expanding away from
the NS from a radius $r_\mathrm{i} = 100$\,km with 
$v_\mathrm{s} = 1000$\,km\,s$^{-1}$ can drag it to a velocity of
$v_\mathrm{ns} \approx 2G\Delta m/(r_\mathrm{i}v_\mathrm{s})
\approx 2700$\,km\,s$^{-1}$. Ejecta asymmetries can thus effectively
mediate a long-lasting pull on the NS.

We do not expect any significant contribution to
the NS kick by anisotropic neutrino emission, because most of
the neutrino energy is radiated from the spherical neutrinospheric
layer \citep{Schecketal06}. Moreover, including the NS motion is
unlikely to change our results, at least statistically
for a larger sample of models. This was concluded by
\citet{Schecketal06} from tests in which the gas surrounding the
NS in the grid center was allowed to move with ($-v_\mathrm{ns}$)
by applying a Galilei transformation. The displacement of the
NS is small compared to the asymmetry scale of the ejecta that 
cause a gravitational acceleration of the NS over seconds. 
\citet{Schecketal06} also observed little influence of the NS motion
on the neutrino emission asymmetry and its associated (very small) 
NS recoil, in contrast to findings by \citet{Fryer04}.

Our 3D simulations were followed much longer beyond the onset of 
the explosion than previous 3D studies by \citet{Fryer04} and 
\citet{FryerYoung07}. This {\em is essential because the 
hydrodynamical NS kicks are no impulsive pre-explosion effect} 
\citep[as, e.g., assumed by][]{JankaMueller94}
{\em but a long-time post-explosion phenomenon}. The global asymmetry 
of the mass-energy distribution in the postshock gas created
before the explosion mainly by SASI activity (as analysed in 
detail by \citet{Schecketal08} for the considered SN core conditions)
is converted to a momentum asymmetry of the ejecta and the NS only 
gradually after the initiation of the blast. Since the 
acceleration continues for seconds, kicks of $\ga$500\,km\,s$^{-1}$
are possible although an extreme dipolar shock deformation
and persistent one-sided accretion as seen in some 2D models of 
\citet{Schecketal06} are absent in the presented 3D results.

The NS spin grows considerably faster than the kick but
saturates also earlier (Fig.~\ref{fig:NSevolution}). Angular
momentum is transferred to the NS by off-center impacts of the
anisotropic accretion downdrafts \citep{Burrowsetal95,SpruitPhinney98}.
When the accretion ends, the exterior
gas separates from the compact remnant. Different from mediating
an acceleration by their gravitational drag, the anisotropic 
ejecta then cannot exert any torque on a spherical, point-like
compact remnant. Therefore the NS spins remain fairly slow, because
in spite of an $\ell = 1,\,m = 1$ density asymmetry we do not find
the strong, ordered rotational gas flows near the accreting PNS as
seen by \citet{BlondinMezzacappa07}. Possibly the investigated
SN core conditions are not favorable for the growth of rotating
spiral modes \citep[see, e.g.,][]{Fernandez10} or
their growth timescale is longer than the explosion 
timescales in our models.

\section{Conclusions}

Our 3D results support the NS recoil scenario proposed on the 
basis of 2D models by \citet{Schecketal04,Schecketal06}. The still 
small set of 3D simulations for two 15$\,M_\odot$ stars yields
kicks between $\sim$50\,km\,s$^{-1}$ and $\sim$500\,km\,s$^{-1}$ 
in the range
of the measured space velocities of young pulsars. Since even
for our most extreme model the explosion asymmetry of the SN is
still fairly small, NS acceleration well beyond 500\,km\,s$^{-1}$
seems possible. But since the kick mechanism is a quasi-chaotic and
stochastic phenomenon, rare cases with more than 1000\,km\,s$^{-1}$
as in 2D \citep{Schecketal06} will require many more model
runs, varying the initial random seed perturbations as well as
the stellar progenitor and explosion energy.

This will also be necessary for statistically significant 
conclusions on the relative orientations of NS spins and kicks. 
Considering nonrotating stars we find spin periods of 
$\sim$500--1000\,ms, but do not see reasons for a
spin-kick alignment. For a closer assessment of this question,
however, also the investigation of progenitors with rotation
is necessary. Core rotation due to the angular momentum
of the progenitor (or established by a first strong thrust) might 
impose a predefined direction and lead to rotational averaging
of the following NS acceleration so that a preferentially aligned
spin-kick distribution emerges instead of random distributions 
of the direction vectors \citep{Wangetal07,Wangetal06,NgRomani07}.
Moreover, even a modest amount of angular momentum can lead to 
higher growth rates of corotating, non-axisymmetric SASI (spiral)
modes with potentially important consequences for NS spins 
and kicks \citep{BlondinMezzacappa07,YamasakiFoglizzo08}.

The recoil scenario described here and by 
\citet{Schecketal04,Schecketal06}
predicts the NS escaping opposite to the direction of the 
maximum explosion strength. We thus suspect that the nickel 
production of the SN and the iron concentration in the SN remnant
could be higher in the hemisphere pointing away from the NS motion.

\acknowledgements
DFG grants EXC153, SFB/TR27, and SFB/TR7, and computing
time at the RZG in Garching are acknowledged.

%%%%%%%%%%%%%%%%%%%%%%%%%%%%%%%%%%%%%%%%%%%%%%%%%%%%%%%%%%%%%
%% Bibliographie
%%%%%%%%%%%%%%%%%%%%%%%%%%%%%%%%%%%%%%%%%%%%%%%%%%%%%%%%%%%%%
%\bibliographystyle{apj}
%\bibliography{kick}

\end{document}